\documentstyle[12pt,psfig,rotate]{article}
\begin{document}
\title{Photoabsorption on nuclei\thanks{Work supported by BMBF and GSI Darmstadt}}
\author{M. Effenberger, A. Hombach, S. Teis and U. Mosel\\
Institut f\"ur Theoretische Physik, Universit\"at Giessen\\
Heinrich-Buff-Ring 16, D-35392 Giessen\\
UGI-96-13}
\date{}
\maketitle
\begin{abstract}
We calculate the total photoabsorption cross section on nuclei 
in the energy range from 300 MeV to 1 GeV 
within the framework of
a semi-classical phase space model. Besides medium mo\-di\-fi\-ca\-tions
like Fermi motion and Pauli blocking we focus on the collision
broadening of the involved resonances. The resonance contributions
to the elementary cross section are fixed by fits to partial  
wave amplitudes of pion photoproduction. The cross sections for
$NR \to NN$, needed for the calculation of collision broadening,
are obtained by detailed balance from a fit to $NN \to NN\pi$ cross
sections. We show that a reasonable collision broadening is not
able to explain the experimentally observed disappearance of the
$D_{13}$(1520)-resonance in the photoabsorption cross section on nuclei.
\end{abstract}
\bigskip
PACS numbers: 25.20.-x
\\{\it Keywords:} total photonuclear cross section, collision broadening
\section{Introduction}
The total nuclear photoabsorption cross section in the
first and second nucleon resonance region which was recently
measured in Mainz \cite{frommhold} and Frascati \cite{bianchi,bianchi2} 
shows clear medium modifications compared to the elementary 
cross section on the proton and deuteron \cite{daphne96,armstrongproton,
armstrongdeut}. One observes a strong broadening of the $\Delta$-peak and
the disappearance of the higher resonances $D_{13}$ and $F_{15}$ while 
the total cross section per nucleon is almost independent of the mass
of the nucleus.
\par In the region of the $\Delta$-resonance there are calculations within
the $\Delta$-hole model \cite{carrascoreal,moniz} that are able to 
describe the experimental data up to 500 MeV. For higher energies there
have been attempts \cite{bianchi2,alberico,kondr94} to deduce the collision
broadening of the resonances from the experimental data by fitting
the total photonuclear cross section; these studies lead to significantly
increased resonance widths in nuclei.
\par We present now a consistent 
calculation of the photon-nucleus reaction over the whole energy range
from 300 MeV to 1 GeV within
the framework of a semi-classical phase space model.
The microscopic input to our model calculation is identical to that used
in the BUU transport model \cite{ber84,cass90}
which has been very successfully applied to the description of
heavy-ion collisions up to bombarding energies of 2 GeV/A \cite{teis},
pion-nucleus reactions \cite{engel93} and also photoproduction of pions
and etas \cite{hombach}. 
Our calculation is based on the assumption that the total photonuclear
cross section is the incoherent sum of contributions from all nucleons
where we neglect possible shadowing effects.
Besides more or less trivial medium modifications
like Fermi motion and Pauli blocking we investigate the effect of
collision broadening for the involved resonances. This may lead to a 
better understanding of the behaviour of nucleon resonances in nuclear
matter.
\par In section \ref{model} we start with the presentation of the cross
sections that we use for the interaction of resonances with nucleons.
The results for the collision widths induced by these cross sections are
shown in section \ref{collision}. The elementary photoabsorption
cross section on the nucleon is discussed in section \ref{ele}.
Finally the results for the total photoabsorption
cross section on nuclei are presented in section \ref{totcross}.
\section{Resonance-nucleon interactions}
\label{model}
We use here resonance parameterizations that we have recently introduced
for use in the collision term for BUU-calculations
that includes baryonic resonances up to a mass of about 2 GeV.
The used decay widths and cross sections are described in detail in
\cite{teis}.
The one-pion decay width is parameterized according to \cite{cass90}:
\begin{equation}
\label{standardbreite}
\Gamma(q)=\Gamma_0 \left( \frac{q}{q_r} \right)^{2l+1}  
\left( \frac{q_r^2+c^2}{q^2+c^2} \right)^{l+1}
\end{equation}
where
\begin{equation}
c^2=(M_R-M_N-m_{\pi})^2+\frac{\Gamma_0^2}{4} \quad.
\end{equation}
Here, $q$ and $q_r$ denote the cms pion momenta of resonances with mass $M$
and $M_R$ respectively where $M_R$ is the mass at the pole of the resonance.
$l$ is the angular momentum of the $\pi N$-system, $M_N$ the nucleon mass
and $m_{\pi}$ the pion mass. $\Gamma_0$ is the decay width at the pole of the
resonance. The used resonance parameters can be found in \cite{teis}.
\par For the $\Delta$-resonance we use the Moniz-parameterization \cite{moniz}:
\begin{equation}
\label{moniz2}
\Gamma_{\Delta}(q)=\Gamma_0 \frac{M_{\Delta}}{M} \left( \frac{q}{q_r} \right)^{3}  
\left( \frac{q_r^2+c^2}{q^2+c^2} \right)^{2}
\quad \mbox{with} \; c=0.3 \, {\rm GeV} \quad,
\end{equation}
because this parameterization gives a good description of the $P_{33}$-multipole
of elastic $\pi N$-scattering within a Breit-Wigner approximation
\cite{arndtpin}.
\par The two-pion decay is parameterized as a two-step process: 
\[ R \to r \, a \to N \, \pi \, \pi \]
where $r$ stands for a baryonic or mesonic resonance and $a$ is a pion or
nucleon. The two-pion decay width is then given by a phase space weighted
integral over the mass distribution of the intermediate resonance:
\begin{eqnarray}
\label{zweipionenbreitef}
\Gamma_{R \to r\,a}(M)&=& \frac{k}{M} \int_0^{M-m_a} d\mu \, p_f \,\frac{2}{\pi} 
\frac{\mu^2\, \Gamma_{r,tot}(\mu)}{(\mu^2-m_r^2)^2+\mu^2\,\Gamma_{r,tot}^2(\mu)} 
\nonumber
\\
&&
\times \frac{\left(M_R-M_N-2m_{\pi}\right)^2+c^2}
{\left(M-M_N-2m_{\pi}\right)^2+c^2} 
\quad \mbox{with} \quad c=0.3\, {\rm GeV}\quad, 
\end{eqnarray}
where $p_f$ denotes the cms momentum of $a$ and $r$ and $\mu$ the mass of the
\linebreak
intermediate resonance. The constant
$k$ follows from the knowledge of 
\linebreak
$\Gamma_{R \to r \, a}(M_R)$.
The parameters of the mesons and the parameterizations of their decay
widths can be found in \cite{teis}.
\par The parameterizations of the partial widths of the $N(1535)$ were fitted to
the eta photoproduction cross section on the proton and are described in
\cite{prodpaper}.
\par For the calculation of collision broadening one needs
to know the cross sec\-tions for interactions of resonances with
nucleons. The cross sections for resonance absorption processes
($N\,R \to N\,N$) are obtained by detailed balance from one-pion production in
nucleon-nucleon collisions under the assumption that the main contribution to 
$N\,N \to N\,N\,\pi$ comes from two-step processes
$N\,N \to N\,R \to N\,N\,\pi$. The contribution
of the $\Delta(1232)$-resonance is adopted from \cite{dimit} where
we have only replaced the parameterization of the $\Delta$-width (equation
(\ref{moniz2})).
This cross section for $N\,N \to N \,\Delta(1232)$
describes experimental massdifferential cross sections reasonably well 
\cite{dimit,teis}.
\par
\begin{figure}[t]
\centerline{
\psfig{figure=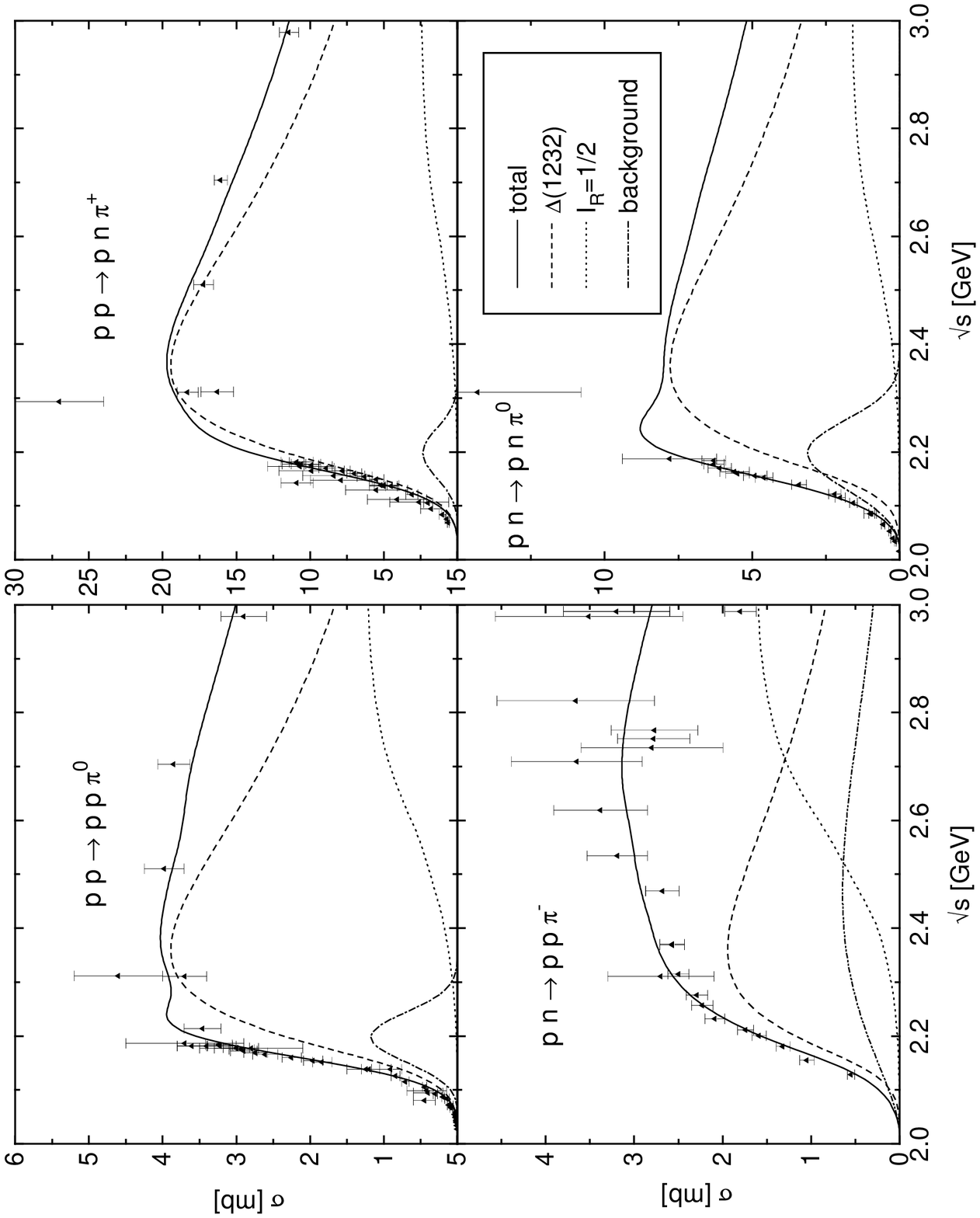,width=11.5cm}}
\caption{Resonance and background contributions to one-pion production
cross sections in nucleon-nucleon collisions. The experimental data are
taken from \protect\cite{landolt2}.}
\label{onepiontot}
\end{figure}

 For the higher resonances massdifferential data are not available. Therefore 
we make the ansatz that
the matrix element for $N\,N \to N\,R$ is constant and equal for all higher resonances.
This matrix element is obtained by fitting
total $N\, N \to N\,N\,\pi$ cross sections. In order to get a better
description of the threshold behaviour a background
term is added. From figure \ref{onepiontot} one sees that the contributions coming 
from the higher resonances are small compared
to that from the $\Delta(1232)$.
\par For the resonance absorption process $N\, R \to N\, N$ we obtain the
cross sec\-tion:
\begin{equation}
\sigma_{N\,R \to N N}=S \, I \, \frac{2}{2J_R+1} \,
\frac{p_f\,\left| \mbox{$\cal M$}_{NN \to N R} \right|^2}{16\pi \, p_i\,s}
\quad,
\end{equation}
where $J_R$ denotes the spin of the resonance, $p_i$ and $p_f$ are the cms
momenta of the incoming and outgoing particles respectively and $\sqrt{s}$
is the total energy in the cms. The isospin factors $I$ result from an
incoherent sum over all contributing isospin channels under the
assumption that the absolute value of the matrix element
$\mbox{$\cal M$}_{NN \to N R}$ does not depend on the total isospin. 
They are given in table \ref{isotable}. The factor $S$ reads:
\[ S=\left\{\begin{array}{r@{\quad , \quad}l}   
\frac{1}{2}& \mbox{identical particles in the final state} \\ 
1&\mbox{else.} \end{array} \right.   \]
\par
The numerical value for the matrix element is fitted to the cross sections
for $N\, N \to N\, N\, \pi$ resulting in:
\begin{equation}
\frac{\left| \mbox{$\cal M$}_{NN \to N R} \right|^2}{16\pi}=
18 \,{\rm mb \, GeV^2} \quad.
\end{equation}
\begin{table}[t]
\centerline{
\begin{tabular}{|ll|ll|c|} 
\hline
 & & & &I\\
\hline
$N^+$ & $N^+$ & $N^+$ & $N^+$ & 1\\
\hline
$N^+$ & $N^0$ & $N^+$ & $N^0$ & 1/2\\
\hline
$N^+$ & $N^+$ & $N^0$ & $\Delta^{++}$ & 3/4\\
\hline
$N^+$ & $N^+$ & $N^+$ & $\Delta^+$ & 1/4\\
\hline
$N^+$ & $N^0$ & $N^+$ & $\Delta^0$ & 1/4\\
\hline
$N^+$ & $\Delta^{++}$ & $N^+$ & $\Delta^{++}$ & 1\\
\hline
$N^+$ & $\Delta^+$ & $N^0$ & $\Delta^{++}$ & 3/8\\
\hline
$N^+$ & $\Delta^+$ & $N^+$ & $\Delta^+$ & 5/8\\
\hline
$N^+$ & $\Delta^0$ & $N^+$ & $\Delta^0$ & 1/2\\
\hline
$N^+$ & $\Delta^-$ & $N^+$ & $\Delta^-$ & 5/8\\
\hline
\end{tabular}}
\caption{Isospin coefficients for baryon-baryon collisions. $N$ stands for any
isospin-$\frac{1}{2}$ particle, $\Delta$ for any isospin-$\frac{3}{2}$ one.}
\label{isotable}
\end{table}
\par
The matrix elements for the $N(1535)$-resonance are fitted to
$N\, N \to N\, N\, \eta$ cross sections. This gives \cite{teis}:
\begin{equation}
\frac{\left|\mbox{$\cal M$}_{p\,p\to p\,N^+(1535)} \right|^2}{16\pi}
=8 {\rm \,mb \, GeV^2}
\end{equation}
\begin{equation}
\mbox{and}\quad \frac{\left|\mbox{$\cal M$}_{p\,n\to n\,N^+(1535)} \right|^2}{16\pi}
={\rm 20 \,mb \, GeV^2} \quad.
\end{equation}
\par
\begin{figure}[t]
\centerline{
\rotate[l]{\psfig{figure=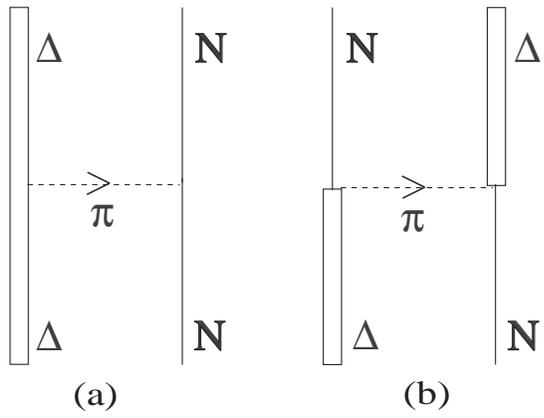,height=8cm}}}
\caption{Diagrams to $\Delta \,N \to \Delta \,N$.}
\label{diagndnd}
\end{figure}
\begin{figure}[t]
\centerline{
\psfig{figure=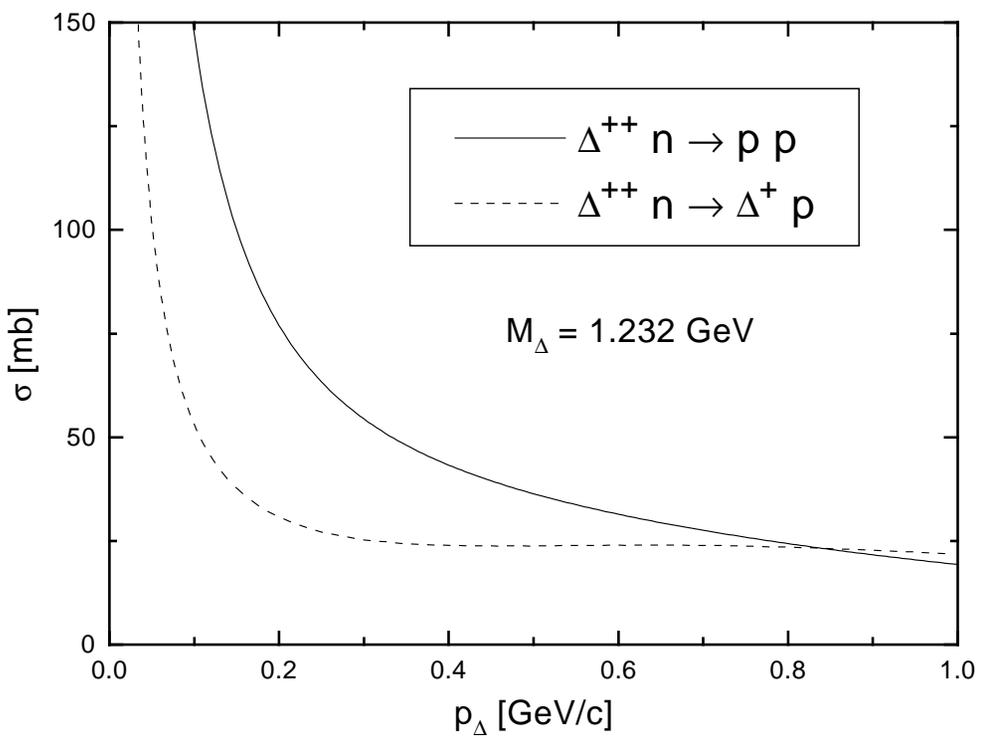,width=11.5cm}}
\caption{Cross sections for $\Delta$-nucleon collisions. $p_{\Delta}$ is
the $\Delta$-momentum in the rest frame of the nucleon.}
\label{ndnd}
\end{figure}

The cross section for $N\,\Delta(1232) \to N\,\Delta(1232)$ is estimated
by the cal\-cu\-la\-tion of the diagram shown in figure \ref{diagndnd} (a)
alone, because the process in figure \ref{diagndnd} (b) with an intermediate
on-shell pion is already included in our transport model via
the two-step process. In figure \ref{ndnd} the resulting
cross section for the one-step process is depicted together with the cross
section for $\Delta(1232)\, N \to N \, N$.  
\par The matrix elements for $N\,R \to N\,R$ (the same resonance in in- and outgoing 
channel) are estimated by using a parameterization of the
elastic nucleon-nucleon scattering cross section \cite{cugnon}:
\begin{equation}
\sigma_{NN \to NN}=\left(\frac{35}{1+\frac{\sqrt{s}-2M_N}{{\rm GeV}}}+
20\right)\,{\rm mb} \quad.
\end{equation}
The assumption of an isotropic angular dependence gives the corresponding
squared matrix element:
\begin{equation}
\left| \mbox{$\cal M$}_{NN \to NN} \right|^2=16\,\pi\,s\left(
\frac{35}{1+\frac{\sqrt{s}-2M_N}{{\rm GeV}}}+20\right)\,{\rm mb} \quad.
\end{equation}
This matrix element is now used for the $N\,R\to N\,R$ cross section: 
\begin{equation}
\sigma_{NR \to NR} = I\,\frac{\left|\mbox{$\cal M$}_{NN \to NN}\right|^2}{16\,\pi\,p_i\,s} 
\int \, d\mu \,p_f \frac{2}{\pi} \,
\frac{\mu^2 \, \Gamma_R(\mu)}{\left( \mu^2-M_R^2 \right)^2+\mu^2\,\Gamma_R^2(\mu)}
\quad,
\end{equation}
where $I$ is again the isospin coefficient from table \ref{isotable} and $\mu$
denotes the mass of the outgoing resonance.
For processes with
a change of the resonance we use the same matrix element as for
$N\,R \to N\,N$. This gives:
\begin{equation}
\sigma_{NR \to NR^{\prime} \atop R \neq R^{\prime}} = I\,
\frac{2}{2J_R+1}\,
\frac{\left|\mbox{$\cal M$}_{NN \to NR}\right|^2}{16\,\pi\,p_i\,s} 
\int \, d\mu \,p_f \frac{2}{\pi} \,
\frac{\mu^2 \, \Gamma_{R^{\prime}}(\mu)}{\left( \mu^2-M_{R^{\prime}}^2 \right)^2+
\mu^2\,\Gamma_{R^{\prime}}^2(\mu)} \, .
\end{equation}
\section{Collision broadening}
\label{collision}
The basic concepts of collision broadening of resonances in nuclei are
described in \cite{kondr94}. We
have used the cross sections for the interaction of resonances
with nucleons which were described in the preceding section
to calculate the collision widths of the resonances that are
important for photonuclear reactions. These are the $P_{33}(1232)$, the $D_{13}(1520)$
and the $F_{15}(1680)$. The $S_{11}(1535)$ is important for the calculation of
etaproduction.
\par Since nucleon final states coming from spontaneous decays of resonances can
be Pauli blocked in the nuclear medium there is also a reduction of the
width. The total in-medium width is therefore:
\begin{equation}
\Gamma_{tot}^{med}=\Gamma_{spon}^{med}+\Gamma_{coll}^{med}\quad,
\end{equation}
where $\Gamma_{spon}^{med}$ stands for the sum of the one-pion-, two-pion- and
eta-width. 
\par The in-medium widths are calculated in nuclear matter and then applied to finite
nuclei by means of a local density approximation. The collision width coming from
the process $N\,R \to N\,N$ for a resonance with mass $M_R$ and momentum
$p_R$ at nucleon density $\rho$ is thus given by:
\begin{equation}
\label{collwidth}
\Gamma_{N_1R \to N_2N_3}(M_R,p_R,\rho)=4\,\gamma\,\int_0^{p_F} \frac{d^3p_{N1}}{(2 \pi)^3}
\,v_r\,\int d\Omega\, \frac{d\sigma_{NR\to NN}}{d\Omega}\,
{\rm P_{N2}}\,
{\rm P_{N3}}\,{\rm S}
\, ,
\end{equation}
where the relative velocity $v_r$ is     
\begin{equation}
v_r=\sqrt{\frac{\left({\bf p}_R \cdot {\bf p}_{N1} \right)^2-M_R^2\,M_N^2}
{\left( {\bf p}_R \cdot {\bf p}_{N1} \right)^2}} \quad,
\end{equation}
with ${\bf p}_R$ and ${\bf p}_{N1}$ being the 4-momenta of the resonance
and incoming nucleon respectively.
The Pauli blocking function
\begin{equation}
{\rm P_N}=\Theta\left( \left|\vec{p}_{N} \right| - p_F \right)
\end{equation}
takes into account that the momenta of the outgoing nucleons have to be outside of the
Fermi sphere. A shadowing function S cuts off the in-medium cross section
at high values:
\begin{equation}
{\rm S}=\left\{\begin{array}{c@{\quad,\quad}l} 1 & \sigma_{NR,tot} 
\le \sigma_{max} \\ \frac{\displaystyle \sigma_{max}}{\displaystyle \sigma_{NR,tot}}&\sigma_{NR,tot}>
\sigma_{max} \end{array} \right. \quad,
\end{equation}
where $\sigma_{max}$ is 
\begin{equation}
\sigma_{max}=\left( \frac{\rho_0}{\rho} \right )^{\frac{2}{3}} \, 80.4\,
{\rm mb} \quad.
\end{equation}
Here,
\[ \sqrt{\frac{80.4 {\rm mb}}{\pi}}=1.6\,{\rm fm} \]
is approximately equal to the spatial distance of neighboring nucleons at
nucleon density $\rho_0$=0.168/${\rm fm^3}$.
The results reported later are insensitive to an increase of $\sigma_{max}$.
\par The collision width for $N\,R \to N\,R^{\prime}$ scattering is calculated
analogously to equation (\ref{collwidth}) with an additional integration
over the mass distribution of the outgoing resonance since the momentum
of the outgoing nucleon entering the Pauli blocking function depends on
the mass $\mu$ of the outgoing resonance:
\begin{eqnarray}
\Gamma_{N_1R \to N_2R^{\prime}}(M_R,p_R,\rho)&=&4\,\gamma\,\int_0^{p_F} \frac{d^3p_{N1}}
{(2\,\pi)^3}\,v_r\,\int d\Omega
\nonumber
\\ 
&&\times
\int d\mu \sum_{R^{\prime}} 
\frac{d^2\sigma_{NR\to NR^{\prime}}}{d\Omega\,d\mu}\,
{\rm P_{N2}}\,
{\rm S}
\quad.
\label{collwidth2}
\end{eqnarray}
We neglect the selfconsistency in this equation due to the width dependence
of the cross sections on the rhs because the effect of the in-medium
width in equation (\ref{collwidth}) compared to the vacuum width is very
small. Moreover one has to keep in mind that the uncertainties in the
cross sections for $N\,R \to N\,R^{\prime}$ are very large.
\par The in-medium one-pion decay width is obtained by averaging over the
decay angle:
\begin{equation}
\Gamma_{R \to N\pi}(M_R,p_R,\rho)=\frac{1}{2} \int_{-1}^1 d\cos(\theta)
\,\Gamma_{0,N\pi}(M_R)\,{\rm P_N} \quad,
\end{equation}
where we assume an isotropic decay in the rest frame of the resonance.
$\theta$ is the angle between outgoing nucleon momentum and boost axis in
the rest frame of the resonance and $\Gamma_{0,N\pi}$ the vacuum one-pion decay width. The
in-medium eta width is calculated analogously.
\par The two-pion decay width is parameterized as a two-step process with a first decay
into a baryonic or mesonic resonance and a pion or nucleon (section
\ref{model}). For
the decay into a baryonic resonance and a pion we assume no medium
modifications while the decay into a nucleon and a mesonic resonance $r$ is
modified due to possible Pauli blocking of the nucleon final state:
\begin{equation}
\Gamma_{R \to r\,N}(M_R,p_R,\rho)=\frac{1}{2} \int_{-1}^1 d \cos (\theta) \int dm_r
\,\frac{d\Gamma_{0,N\,r}(M_R)}{dm_r}\,{\rm P_N} \quad,
\end{equation}
where $m_r$ denotes the mass of the outgoing mesonic resonance.
\par
\begin{figure}[t]
\centerline{
\rotate[r]{\psfig{figure=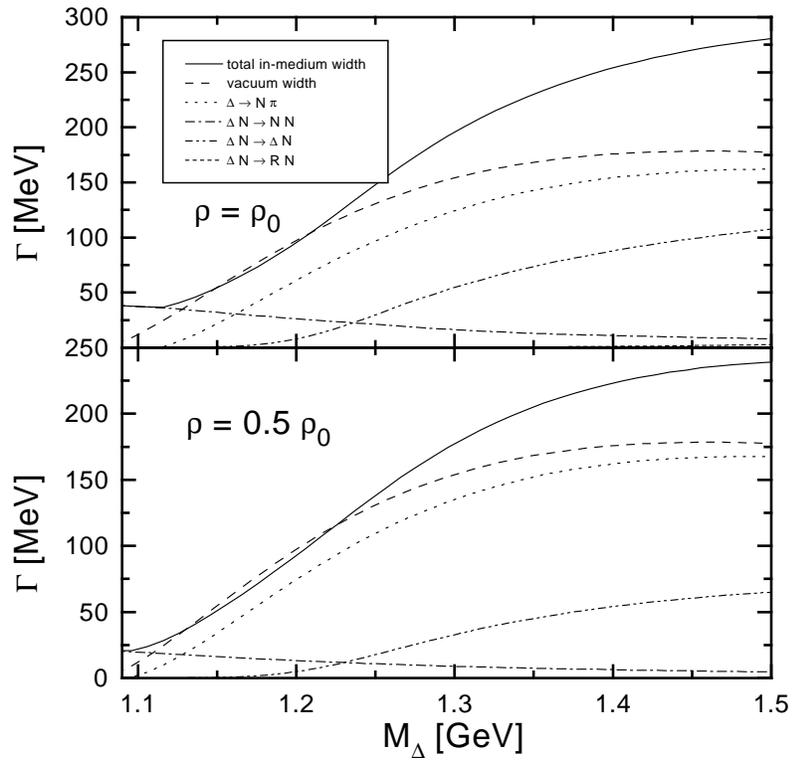,height=11.5cm}}}
\caption{In-medium width of the $\Delta$(1232).}
\label{delwidth}
\end{figure}

 In figure \ref{delwidth} we show the different contributions to the
in-medium width of the $\Delta(1232)$ at nucleon densities $\rho_0$ and
$\rho_0/2$ in isospin symmetric nuclear matter. Here the momentum of the
resonance is related to its mass by the requirement that the resonance was
created by photoabsorption on a free nucleon at rest. For comparison the
vacuum width is also shown.
\par At $\rho_0$ the collision width coming from
resonance absorption $N\,\Delta \to N\, N$ is at the pole of the resonance
($M$=1.232 GeV) about 25 MeV. This partial width decreases with increasing
mass. The contribution from $N\,\Delta \to N\,\Delta$ has - averaged over the
mass distribution - about the same size as the one from $N\,\Delta\to N\,N$.
However, here we get a strong increase of $\Gamma_{N \Delta \to N \Delta}$
with increasing mass because of the
phase space weighted integral over the mass distribution of the outgoing
$\Delta$-resonance (equation (\ref{collwidth2})). 
\par Our calculations of the collision widths for the $\Delta$-resonance are in
agreement with the estimates given by Kondratyuk et al.
(chapter 4 in \cite{kondr94}). The process $N\,\Delta \to N\,R$ is negligible
because the energy of the $N\,\Delta$-system is too low.
\par In the region of the resonance pole the total in-medium width is almost
independent of the nucleon density since collision broadening and Pauli
reduction of the free width nearly compensate. Thus, at the resonance pole
the net broadening
compared to the vacuum width is very small; at about 100 MeV above the pole
the width has grown by about 50 MeV, mainly due to the
$N \, \Delta \to N \, \Delta$ scattering process.
\par
\begin{figure}[t]
\centerline{
\psfig{figure=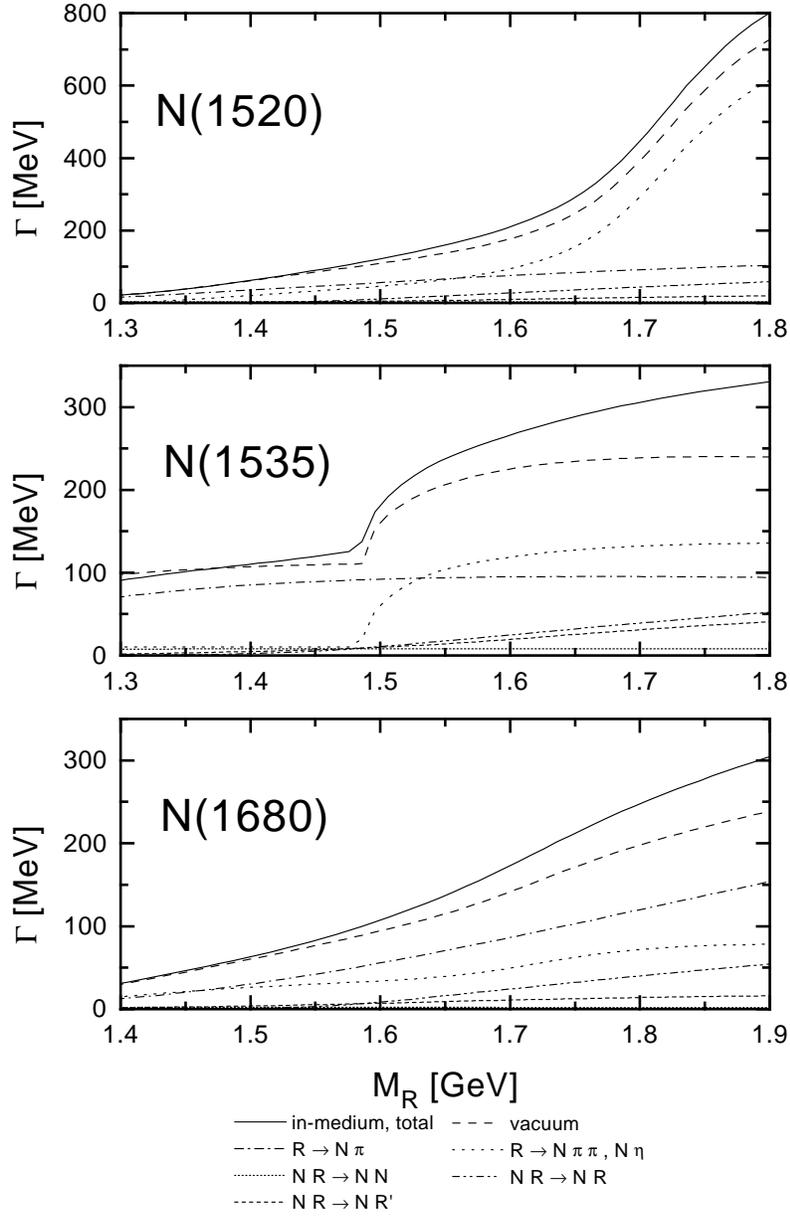,width=11.5cm}}
\caption{In-medium widths of the higher resonances that are relevant for
photonuclear processes.}
\label{hrestot}
\end{figure}

 In figure \ref{hrestot} the in-medium widths of the $N(1520)$, the
$N(1535)$ and the $N(1680)$ are compared with the vacuum widths and split
up into their partial widths. 
The very strong rise of the width of the $N(1520)$ is due to the opening
of the $\rho$-channel which is probably not cut off fast enough.
Ne\-ver\-the\-less, in all cases
the collision widths at the poles of
the resonances are only of the order 20 - 40 MeV which leads to a small net broadening
because the Pauli blocking of the free width is less important than in the
case of the $\Delta$-resonance. Notice that the collision widths of those higher
lying resonances increase less significantly with mass than those of the
$\Delta$.
\par The main contribution to the collision widths comes from the process $N\,R \to N \,R$ for which
we estimated the cross section by adopting the matrix element from elastic
nucleon nucleon scattering. The partial widths of $N\,R \to N\,N$ are
very small ($<$ 10 MeV) as to be expected from figure \ref{onepiontot}. Even
if one dropped the assumption that the matrix element is equal for all higher
resonances and set the matrix element of the $D_{13}(1520)$ to the maximum
possible value which is in line with the $N\,N \to N\,N\,\pi$ data, one
would only get an additional broadening of the $D_{13}(1520)$ of about
30 MeV. Therefore a collision width of 300 MeV for the $D_{13}(1520)$ as
given in \cite{alberico,kondr94} seems to be far from being realistic.
\par In order to check the validity of the local density
approximation used for the application of the discussed in-medium widths
we also performed a calculation of the $\Delta$-width in finite nuclei.
Here we defined the density to be the density at the location of creation
of the resonance. It turned out that there was a difference between this
calculation and the one in nuclear matter only for very low mass $\Delta$'s
because the width of these $\Delta$'s is small enough for them to travel
through a relevant density gradient within their lifetimes. For $\Delta$'s in the
region of the resonance pole the mean free path is only about 0.5 fm
leading to a sensible applicability of the local density approximation.
\par In summary, the effective collisional broadening of the resonances never
amounts to more than 10\% at the resonance pole.
\section{The total photoabsorption cross section on the nucleon}
\label{ele}
In \cite{prodpaper} the decomposition of the total photoabsorption cross
section on the nucleon into the different channels and the resonance
contributions is described in detail. For the one-pion production cross
section we use partial-wave amplitudes as given by \cite{Arndt} and fit
the resonance contributions to these amplitudes because - especially in
the region of the $\Delta$-resonance - interference terms with the background
are quite important. An incoherent de\-com\-po\-si\-tion of the total photoabsorption
cross section into resonance and background contributions as done by
Kondratyuk et al. \cite{kondr94} should not be used if one wants to
investigate possible modifications of the resonance contributions in nuclei. 
\par While the one-pion production cross sections can nicely be decomposed into
Breit-Wigner type resonance contributions and a smooth background, the
structure of the two-pion production cross sections is not described by the
resonance contributions that are induced by the two-pion decay widths of the
resonances \cite{prodpaper}. The difference between the
experimental cross section and the Breit-Wigner type resonance contributions
is treated
as background, where the momenta of the outgoing particles are distributed
according to three-body phase space. Therefore the only medium modification is
the possible Pauli blocking of the outgoing nucleon.
\par
\begin{figure}[t]
\centerline{
\psfig{figure=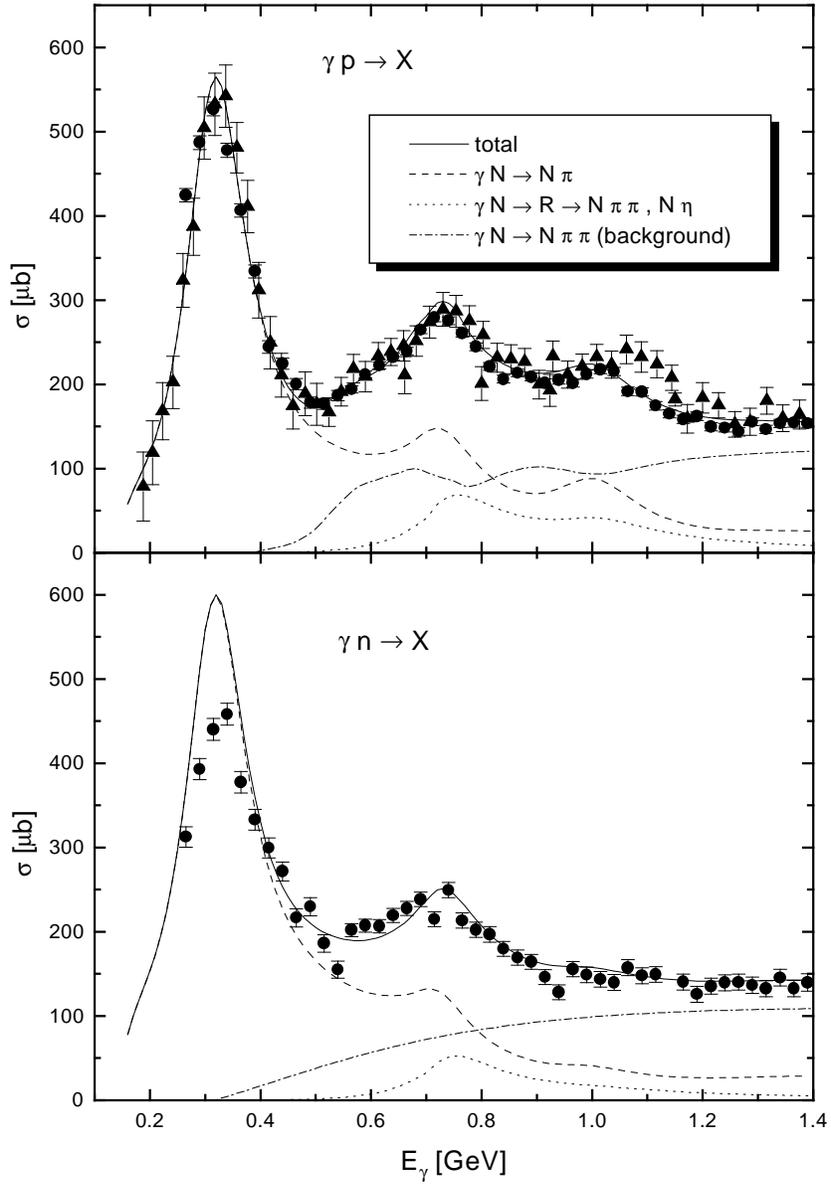,width=11.5cm}}
\caption{Total photoabsorption cross sections on the nucleon. 
The experimental proton data are from \protect\cite{armstrongproton} 
(circles) and \protect\cite{landolt2}
(triangles). The neutron data are taken from \protect\cite{armstrongdeut}.}
\label{gamabs1}
\end{figure}

 In figure \ref{gamabs1} we show the sum of the contributions to the
total photoabsorption cross section on the proton with the experimental data
\cite{armstrongproton,landolt2}. The good agreement indicates that all
important channels are included. For photon energies above 600 MeV the
'background' for $\gamma \,p \to N \, \pi \, \pi$ amounts for about one half
of the total cross section.
\par For the neutron the two-pion background was fitted to the absorption
cross section \cite{prodpaper}. 
In the region of the $\Delta$-resonance the sum of the one-pion cross
sections \cite{Arndt} alone is about 150 $\mu$b larger than the total absorption cross
section pub\-lished by Armstrong et al. \cite{armstrongdeut}. Since
the photoabsorption cross section on the deuteron measured by Armstrong et
al. has recently be confirmed by the DAPHNE collaboration \cite{daphne96} we
assume this
discrepancy to arise from the extraction of the neutron cross section from
the deuteron data. 
\par Unfortunately there is yet no calculation of the
photoabsorption on the deuteron in the energy range of interest. 
However, there is a recent calculation of quasifree pion photoproduction
on the deuteron in the $\Delta$-region within the spectator nucleon model
\cite{schmidt96}. Here, the cross section for $\gamma \, d \to N\, N\, \pi$ 
at the $\Delta$-peak is about 1100 $\mu$b. This result is at variance
with the measured total photoabsorption cross section on the 
deuteron being only about 900 $\mu$b \cite{daphne96,armstrongdeut}.
\subsection{Medium modifications}
We use the following medium modifications for the elementary photon-nuc\-le\-on
cross section:
\begin{itemize}
\item The vacuum width appearing in the resonance propagators
is replaced by the in-medium width as calculated
in section \ref{collision}.
\item The collision width gives a Breit-Wigner type contribution to the
absorption cross section \cite{prodpaper}.
\item For the $\Delta$-resonance the difference between nucleon and
$\Delta$-potential causes a real part
of the self energy $\Pi$ to be used in the resonance propagator:
\begin{equation}
{\rm Re}\,\Pi=2\,E_{\Delta}\,\left( U_N-U_{\Delta} \right) \quad.
\end{equation}
\item Nucleon final states can be Pauli blocked. 
\end{itemize}
\section{Results}
\label{totcross}
We now combine the ingredients described in the preceding sections in a
calculation of the total photoabsorption cross section on nuclei using the
local Thomas-Fermi model with realistic density distributions.
Neglecting shadowing effects the total cross section on a
nucleus within our model then reads:
\begin{eqnarray}
\sigma_{abs}&=&\int d^3r \int^{p_F}4\,\frac{d^3p_N}{(2\pi)^3}
\frac{k_{\gamma N}}{k_{\gamma A}} \frac{M_N}{E_N}
\left\{
\int \left[\frac{Z}{A} 
\left(
\frac{d\sigma^{med}_{\gamma p \to N\pi}}{d\Omega}+
\frac{d\sigma^{med}_{\gamma p \to p\eta}}{d\Omega}
\right)
\right. \right.
\nonumber
\\
&&
\left.
+\frac{A-Z}{A} 
\left(
\frac{d\sigma^{med}_{\gamma n \to N\pi}}{d\Omega}
+\frac{d\sigma^{med}_{\gamma n \to n\eta}}{d\Omega} 
\right)
\right]d\Omega
\nonumber
\\
&&
+
\left.
\int \left[\frac{Z}{A} \frac{d\sigma^{med}_{\gamma p \to N\pi\pi}}{d\Omega^{\prime}\,dp_N^{\prime}}+
\frac{A-Z}{A} \frac{d\sigma^{med}_{\gamma n \to N\pi\pi}}{d\Omega^{\prime}\,dp_N^{\prime}} 
\right]d\Omega^{\prime}\,dp_N^{\prime}
\right\}
\quad,  
\end{eqnarray}
where $p_F[\rho(r)]$ denotes the local Fermi momentum \cite{prodpaper},
$k_{\gamma N}$ is the photon energy in the rest frame of the nucleon with
momentum $\vec{p}_N$ and $k_{\gamma A}$ the photon momentum in the rest
frame of the nucleus. The factor
\[ \frac{k_{\gamma N}}{k_{\gamma A}} \frac{M_N}{E_N} \]
results from the Lorentz transformation of the cross section \cite{carras92}.
\par
\begin{figure}[t]
\centerline{
\psfig{figure=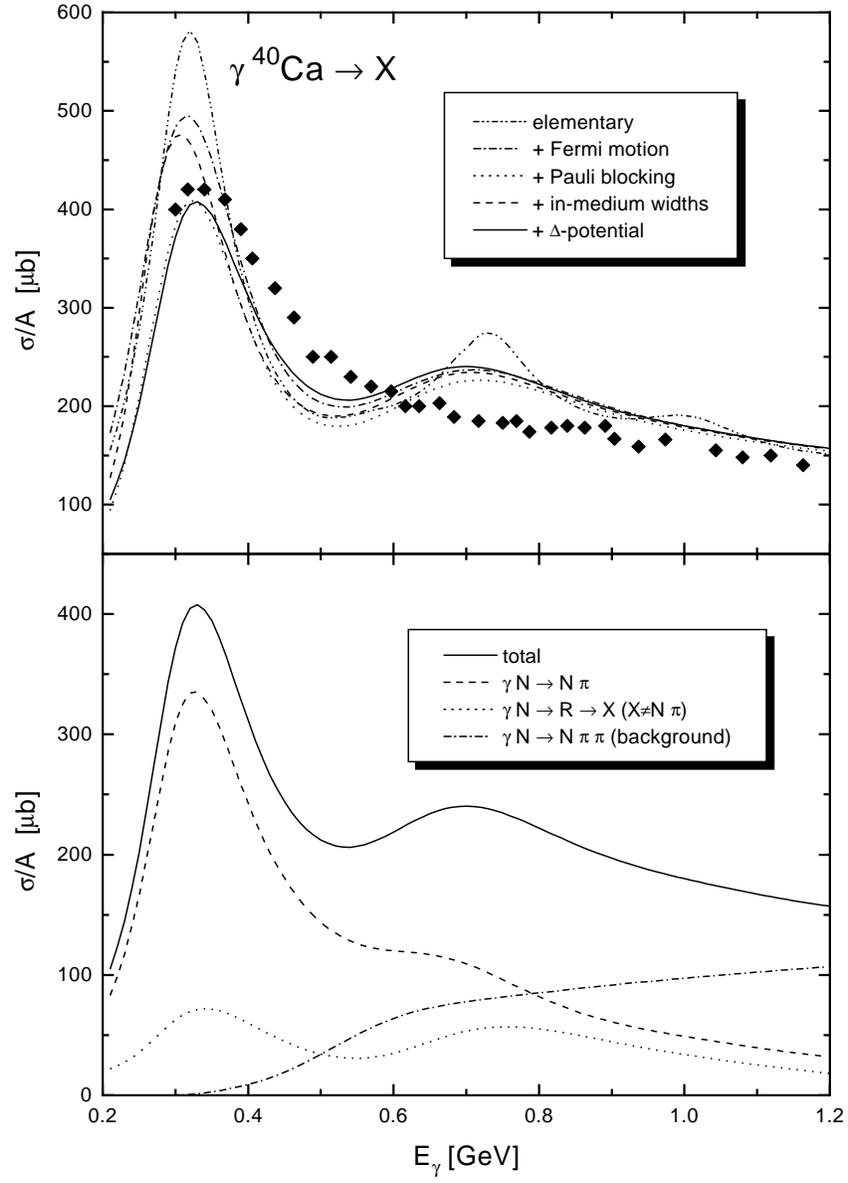,width=11.5cm}}
\caption{Photoabsorption cross section on $^{40}$Ca. The experimental data
are obtained by an average over different nuclei \protect\cite{bianchi}.}
\label{kgamabs}
\end{figure}

 Figure \ref{kgamabs} shows the cross sections on $^{40} {\rm Ca}$ that
result from successive application of the medium modifications. Fermi
motion alone leads to a damping of the $\Delta$-peak by about 100 $\mu$b per
nucleon. The structure in the region of the $D_{13}$-resonance is washed
out but does not disappear. The peak of the $F_{15}$-resonance vanishes.
Pauli blocking further decreases the $\Delta$-peak by about 100 $\mu$b with
the reduction of the cross section getting smaller at higher energies.
\par Using the in-medium widths for the resonances the $\Delta$-peak is shifted
to smaller energies and increased by about 70 $\mu$b per nucleon. The reason
is given by the comparison of the $\Delta$-in-medium width with the vacuum
width in figure \ref{delwidth}. Close to the resonance pole in-medium width
and vacuum width are almost equal leading to an increase of the absorption
cross section due to the contribution from the collision width. Because of
the strong increase of the in-medium width with increasing $\Delta$-mass
the $\Delta$-contribution is decreased at higher energies. This cancels the
contribution from the collision width at photon energies of about 400 MeV.
\par In these calculations the $\Delta$'s experience the same potential as the
nucleons. If we apply a $\Delta$-potential of $U_{\Delta}=-30 \,\rho/\rho_0\,
{\rm MeV}$ the $\Delta$-peak is
shifted to higher energies and decreased by about 70 $\mu$b per nucleon
because the $\Delta$-width increases strongly with increasing mass and the
$\Delta$-peak is proportional to $\frac{1}{\Gamma}$. Since the position
of the $\Delta$-peak becomes density dependent the integration over the
volume of the nucleus leads to a further smearing out. It is im\-por\-tant
to note that a different parameterization of the $\Delta$-width, for
example a constant width, is not in line with the $P_{33}$-multipole
of elastic $\pi N$-scattering \cite{arndtpin}. Therefore the $\Delta$-peak
has to be reduced if shifted to higher photon energies. Compared to the
experimental data \cite{bianchi} we see from figure \ref{kgamabs} that we
underestimate the cross section in the high mass $\Delta$-region.
\par The very different $\Delta$-potential of
$U_{\Delta}=0 \,{\rm MeV}$
leads to a reduction of the $\Delta$-peak by about 50 $\mu$b and thus does not give a
better description of the experimental data.
\par Using a larger cross section for $\gamma \,n \to n \,\pi^0$ according
to an alternative in \cite{Kruschepi0} gives only a little improvement in
the description of the experimental data.
\par In the region of the $\Delta$-resonance a better description of the
experimental cross section is only possible by an extension of our model.
The missing strength indicates that we have to include further reaction
mechanisms.
\par
\begin{figure}[t]
\centerline{
\rotate[l]{\psfig{figure=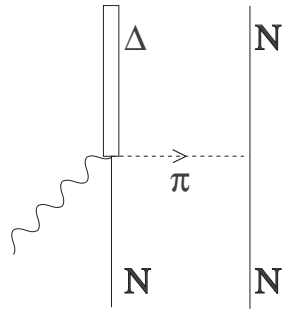,height=6cm}}}
\caption{Example of a diagram contributing to
$\gamma \, N \, N \to N \, \Delta(1232)$.}
\label{2absd}
\end{figure}

\begin{figure}[t]
\centerline{
\psfig{figure=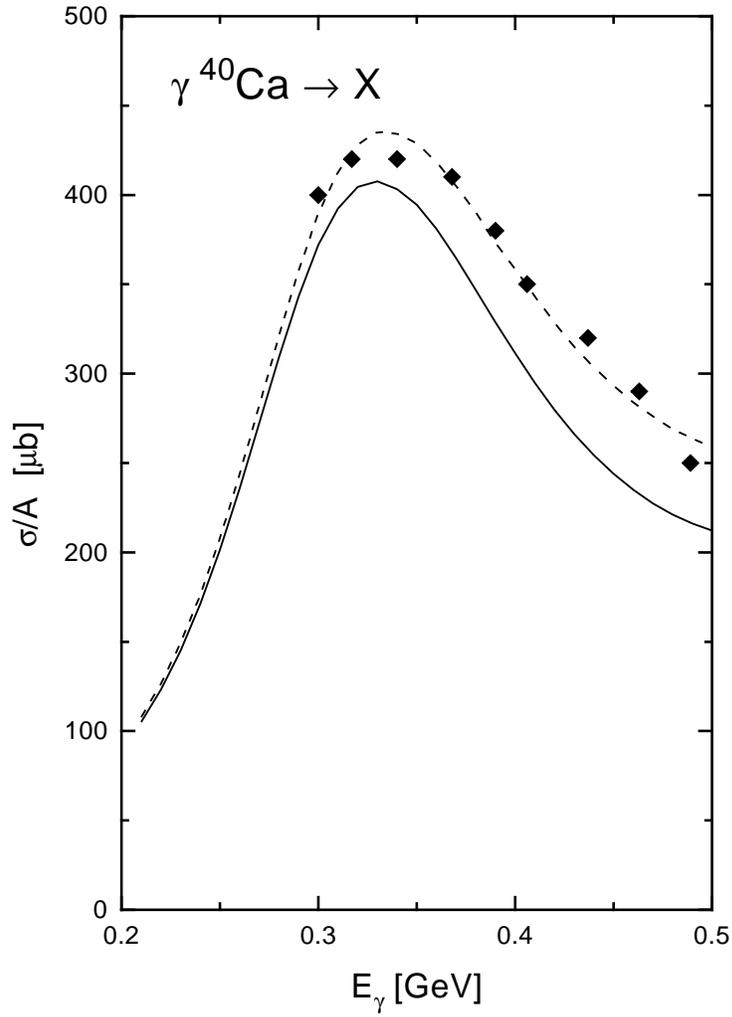,width=11.5cm}}
\caption{Total photoabsorption cross section on $^{40}$Ca with the additional
two-body absorption mechanism $\gamma \, N \, N \to N \, \Delta(1232)$ of
figure \ref{2absd} (dashed line). The
solid line represents the previous cross section.}
\label{2abst}
\end{figure}

 In the framework of the $\Delta$-hole model there are calculations
of the photoabsorption cross section up to energies of 500 MeV
in \cite{carrascoreal} and up to 400 MeV in \cite{moniz}. The results of both
calculations differ by about 10\%. The calculation in
\cite{carrascoreal} gives a very good description of the experimental data.
The authors claim that a two-body absorption process as depicted in figure
\ref{2absd} plays an important role in the energy range between
400 and 500 MeV. Our calculation of this diagram gives a similar result.
Adding this contribution to the cross section described above leads to
a good description of the experimental data (figure \ref{2abst}).
Of course, just adding this one amplitude does not represent a gauge
invariant calculation, but, in agreement with \cite{carrascoreal}, it
gives the right direction.
\par In the region of the higher resonances the effect of the in-medium widths
is small since the collision widths are small as discussed in section
\ref{collision}. Here the collision widths essentially only compensate
the Pauli reduction.
At a photon energy of 700 MeV we still obtain a resonance-like
structure and overestimate the experimental cross section by about
25\%.
\par The main reason for the observed insensitivity of the $N(1520)$-resonance
region to the model ingredients lies thus in the fact that the collisional
widths are rather small compared to the decay width. In a consistent
treatment one also has to take into account the collision broadening
of the nucleon ground state. This effect can be estimated from the
imaginary parts of the selfenergies of nucleons in nuclear matter and amounts,
depending on the nucleon momentum, to 0 - 60 MeV broadening with an
average width being less than 10 MeV \cite{fred}.
In order to mimic this effect and other possible in-medium modifications
we have used the ansatz:
\begin{equation}
\Gamma_{coll}^*=\Gamma_{coll} \left(\alpha+\beta\,\frac{\rho}{\rho_0} \right)
\end{equation}
The $\Delta$-peak is not only reduced with increasing
$\alpha$ or $\beta$ but also shifted to smaller photon energies since the collision
width from $\Delta\, N \to N\, N$ does not - in contrast to the free width -
vanish for small $\Delta$-masses. For reasonable values of $\alpha$ and
$\beta$ the higher resonances are not affected.
\par In figure \ref{kgamabs} (lower part) we also show the different contributions to the
total cross section. Here we see that the rise of the cross section between
550 and 700 MeV is not only caused by the excitation of the $D_{13}$-resonance
but also to a large extent by the opening of the two-pion background channel. The contribution
of the one-pion channel shows almost no resonant structure in this energy
regime.
\par Alberico et al. \cite{alberico} and Kondratyuk et al. \cite{kondr94}
have performed phenomenological fits to the photoabsorption cross section on
nuclei in order to gain information about masses and widths of
nucleon resonances in nuclei.
The main result of these studies was the need for significantly increased
resonance widths, up to a factor of 4 for the $D_{13}$-resonance to explain
the observed absence of the higher lying resonances. If we use a collision
width of about 300 MeV for the $D_{13}$ then also in our calculation the
resonant structure in the cross section disappears.
However, as discussed
in section \ref{collision}, a collision width of 300 MeV for the
$D_{13}$-resonance is hard to justify.
Furthermore, in \cite{alberico} the neglect
of background terms as well as the assumption that the elementary cross
section on the neutron is equal to the one on the proton has led 
these authors to an unsatisfactory
determination of the resonance couplings to the $N\gamma$-system in the
vacuum. 
In \cite{kondr94}, furthermore, the cross section of Armstrong et al. for
$\gamma \,n \to X$ was used; as discussed in section \ref{ele} this cross
section is probably incorrect. The fit in the region of the $\Delta$-resonance
gave a good result in \cite{kondr94} only because a constant $\Delta$-width was used and
Pauli blocking for the background was neglected.
The fit of Kondratyuk et al. has been improved in a recent work by
Bianchi et al. \cite{bianchi2}. However, the 'background' as well as the 
neutron cross section are still treated in an unsatisfactory way. 
\par While the collisional broadening can, according to our results, not explain
the observed disappearance of the higher resonances, there are other
in-medium effects that still have to be explored. For example, there is
the possibility that the width of the $D_{13}$-resonance is
increased by a strong medium modification of the free width, for example
caused by the strong coupling to the $N\, \rho$-channel and a downward mass shift of
the $\rho$-meson in the nuclear medium \cite{rho}. This may lead to the disappearance
of the structure in the region of the $D_{13}$-resonance. Another
possibility is a strong medium modification of the elementary
$\gamma \, N \to N \, \pi \, \pi$ cross section in the nuclear medium or
a medium effect on the background amplitudes.
\par An understanding of the disappearance of the $D_{13}$-resonance in the
photoabsorption cross section might thus
be possible by a comparison between theory and experiment with respect
to more exclusive reaction channels \cite{prodpaper}; for example, a
measurement of two-pion photoproduction in nuclei would give information
on the opening of the 2$\pi$-channel in the medium.
\section{Summary and outlook}
We have presented a calculation of the photoabsorption cross section on
nuclei within a local Thomas-Fermi
framework for photon energies
from 300 MeV to 1 GeV. Starting from a reasonable parameterization of the
free photon nucleon cross section we applied the medium modifications
Fermi motion, Pauli blocking and collision broadening for the involved
nucleon resonances.
\par The calculation of collision broadening required the
knowledge of the interaction of nucleon resonances with nucleons. The cross
section for $N\, R \to N\, N$ was obtained by detailed balance from one-pion
production cross sections in nucleon-nucleon collisions under the assumption
of a simple resonance model. Using this information we also estimated cross
sections for $N\, R \to N\, R^{\prime}$. For the $\Delta (1232)$-resonance
it turned out that collision broadening and reduction of the free width
by Pauli blocking nearly compensate each other resulting in a very small net
broadening. The collision broadening of the higher resonances in our
model is almost negligible.
\par Our calculated photoabsorption cross section fails to describe the
experimentally observed disappearance of the
$D_{13}$-resonance. This might be caused by a strong
broadening of the $D_{13}$-resonance in the nuclear medium due to a strong
coupling
to the $N \rho$-channel and a mass shift of the $\rho$-meson in the
nuclear medium
but also by
a medium modification of the $\gamma \,N \to N \, \pi \, \pi$ process.
\par An experimental measurement of exclusive cross sections would be helpful for
a better understanding of the photon-nucleus reaction and an explanation
of the disappearance of the $D_{13}$-resonance in the total photonuclear
absorption cross section. In particular, the experimental investigation
of the 2$\pi$-channel on nuclei would be very important because of the
opening of this channel in the $N(1520)$-resonance region.
\bigskip
\par We gratefully acknowledge stimulating discussions with Volker Koch.

\newpage
\clearpage
\clearpage
\clearpage
\clearpage
\clearpage
\clearpage
\clearpage
\clearpage
\end{document}